# The Role of GitHub Copilot on Software Development: A Perspective on Productivity, Security, Best Practices and Future Directions

**Suresh Babu Nettur[1*], Shanthi Karpurapu[1*], Unnati Nettur[2], Likhit Sagar Gajja[3], Sravanthy Myneni[1], AND Akhil Dusi[4]**

[1] Independent Researcher, Virginia Beach, VA 23456, USA
[2] Department of Computer Science, Virginia Tech, Blacksburg, VA 24061 USA
[3] Department of Computer Science, BML Munjal University, Haryana 122413 INDIA
[4] Department of Information Systems, Indiana Tech, Indiana USA

Corresponding authors: Shanthi Karpurapu (shanthi.karpurapu@gmail.com), Suresh Babu Nettur (nettursuresh@gmail.com)

[*] Shanthi Karpurapu and Suresh Babu Nettur are co-first authors

**ABSTRACT** GitHub Copilot is transforming software development by automating tasks and boosting productivity through AI driven code generation. In this paper, we conduct a literature survey to synthesize insights on Copilot's impact on productivity and security. We review academic journal databases, industry reports, and official documentation to highlight key findings and challenges. While Copilot accelerates coding and prototyping, concerns over security vulnerabilities and intellectual property risks persist. Drawing from the literature, we provide a perspective on best practices and future directions for responsible AI adoption in software engineering, offering actionable insights for developers and organizations to integrate Copilot effectively while maintaining high standards of quality and security.

## I. INTRODUCTION

The evolution of code generation tools has significantly transformed the software development landscape, enabling developers to automate repetitive tasks, accelerate coding processes, and improve overall productivity. As these tools have advanced, artificial intelligence integration has enabled more sophisticated solutions, with tools such as GitHub Copilot emerging as significant players in the field. Early iterations of code generation tools focused on simplifying template-based code creation, assisting with boilerplate code, and offering solutions for specific programming languages or frameworks. Recently, AI assisted code generation tools leverage models such as Large Language Models (LLMs) to predict, auto complete, and write complex code snippets based on contextual input from the developer. These tools aim to reduce development time, minimize errors, and assist developers by making code suggestions in real time. In recent years, GitHub Copilot has become a leading AI powered code generation tool, integrating seamlessly into developer environments and redefining the concept of collaborative coding with AI.

GitHub Copilot, initially launched in 2021 and developed in collaboration with OpenAI, is an AI powered tool designed to assist developers with code generation. It is powered by OpenAI's Codex model, a version of GPT-3 specifically designed for code generation. Codex leverages deep learning techniques and is trained on a massive data set that includes public GitHub repositories and other open source code. This dataset features 159 gigabytes of Python code from 54 million repositories, enabling Codex to generate accurate and context aware code across various programming languages. GitHub Copilot Chat now operates on OpenAI's GPT-4o and offers early access to OpenAI o1, a model known for excelling in complex reasoning tasks and demonstrating improved performance in benchmark evaluations [1].

GitHub Copilot seamlessly integrates with popular Integrated Development Environments (IDEs) like Visual Studio Code, Visual Studio, Neovim, and JetBrains. It offers real time code suggestions with auto completion and can generate entire functions based on natural language descriptions or existing code snippets. What distinguishes GitHub Copilot from traditional code completion tools is its capability to understand broader coding contexts and predict the next logical steps of the development process. This feature is especially beneficial for developers who work across multiple programming languages and technologies, as Copilot supports various languages, including but not limited to Python, Java Script, TypeScript, Ruby, Angular, and Go. As developers use Copilot, it continually refines its suggestions based on user input and project context, providing code snippets tailored to specific coding styles. This interactive learning method transforms Copilot into an AI partner in software development. The key features of GitHub Copilot [2] are listed in Table 1.

| Feature | Description |
|---|---|
| Code Completion | GitHub Copilot provides auto complete style suggestions in supported IDEs (e.g., Visual Studio Code, Visual Studio, JetBrains, Azure Data Studio, Vim/Neovim). Users can select standard commands like /fix (fix the problem in code), /explain (offer detailed explanations of code), /doc (generate documentation for code), and /tests (create tests for the code) for a selected portion of text. Alternatively, users can enter a query to receive tailored suggestions, enabling real time code improvements based on AI driven insights. |
| Copilot Chat | GitHub Copilot provides a chat interface for coding related questions. Users can select standard commands such as /fix, /explain, /doc, or /tests for a selected portion of text. Alternatively, users can enter custom queries to receive Copilot's context aware suggestions and solutions in real time. Users can check log errors, create feature flags, and deploy apps to the cloud (Public Beta). Additional features include "pull request difference analysis," a web search powered by Bing (Public Beta), and the ability to inquire about failed Actions jobs (Public Beta). Users can also obtain answers regarding issues, pull requests, discussions, files, commits, and more [2]. |
| Copilot in the CLI | A chat like interface in the terminal that provides command suggestions or explanations. |
| Pull Request Summaries | AI generated summaries of pull request changes, emphasizing affected files and critical areas for review (Copilot Enterprise only). |
| Text Completion (Beta) | AI powered text completion to quickly and accurately generate pull request descriptions (Copilot Enterprise only). |
| Knowledge Bases | Create and maintain documentation sets to serve as contextual references for conversations with GitHub Copilot. When using Copilot Chat on GitHub.com or in Visual Studio Code, you can select a specific knowledge base to improve the relevance and accuracy of Copilot's responses to your queries. |

TABLE 1: GITHUB COPILOT KEY FEATURES.

Since its launch, GitHub Copilot has experienced rapid adoption within the developer community. Recent statistics show that GitHub Copilot has been integrated into millions of developer environments worldwide, with approximately 14.5 million downloads in Visual Studio, 20 million in Visual Studio Code, and 10 million in JetBrains. Over 400 organizations have adopted GitHub Copilot, according to GitHub's 2023 report, and this number is expected to grow significantly following the launch of GitHub Copilot for Business [3].

The central goal of our research is to advance the application of deep learning and LLMs across software engineering [4][5] and medical diagnostics [6][7]. In this paper, we conducted an extensive literature survey to synthesize key insights on GitHub productivity and security implications. Our approach involved selectively reviewing academic databases, along with examining industry white papers, technical reports, official documentation from sources like GitHub and OpenAI. Rather than aiming for an exhaustive systematic review, we focused on identifying studies most critical from our perspective, analyzing their findings, and distilling key insights. This process enabled us to recognize pressing challenges, evaluate existing solutions, and formulate a perspective on best practices and future directions to effectively address these issues.

## II. Productivity Impacts of GitHub Copilot

GitHub Copilot is well recognized in the software developer community and is known for its ability to enhance productivity by automating various coding tasks. It accelerates rapid prototyping and experimentation, enabling developers to quickly generate code snippets and test new ideas by providing context aware suggestions. GitHub Copilot offers a range of use cases that enhance productivity in the software development process, as summarized in Table 2.

.

| Use case | Description |
|---|---|
| Routine Task Automation | Experienced developers benefit from Copilot by automating repetitive tasks, allowing more time for complex work. Senior developers often manage multiple projects and find that tools like Copilot save time on routine tasks such as writing unit tests or database queries. |
| Learning and Skill Development | Junior developers benefit from Copilot as an interactive tutor, helping them quickly learn unfamiliar programming languages, frameworks, or libraries. By suggesting optimized code and providing immediate feedback, Copilot enhances their coding skills and boosts their confidence. The 'Explain this' feature of Copilot helps junior developers break down and understand complex logic or algorithms easily. |
| Code Refactoring | Copilot greatly assists developers in cleaning up legacy codebases by identifying redundancies and recommending reusable code blocks. This enhances code readability and maintainability, ultimately accelerating future development efforts. |
| Code Review | Developers can benefit from GitHub Copilot during code reviews by receiving recommendations that help maintain high standards of quality and consistency. Copilot's pull request (PR) feature streamlines this process by automatically generating summaries of changes and highlighting key areas that need attention. This allows teams to stay productive while ensuring that quality is not sacrificed, resulting in a more efficient and detailed review process. |
| Test-Driven Development (TDD) | Developers benefit from Copilot by supporting TDD practices in the generation of test cases. This enables them to implement TDD seamlessly and efficiently in the early stages of development, ensuring higher code quality from the outset. |

TABLE 2: GITHUB COPILOT COMMON USE CASES IN SOFTWARE DEVELOPMENT

### A. UNDERSTANDING PRIOR RESEARCH IN CONTEXT

To understand the productivity implications of GitHub Copilot, we have reviewed relevant research papers. Our analysis reveals that several studies indicate significant productivity improvements associated with Copilot. These studies consider a range of factors such as number of developers, experience levels of developers, types of coding tasks performed, programming languages used, complexity of the tasks, and various demographic details. This breadth of analysis allows us to gain a nuanced understanding of Copilot's impact on productivity across different segments of the developer community.

A recent study conducted by Solohubov et al. on GitHub Copilot evaluates its impact on developer productivity, particularly in the context of creating CRUD operations using the Dart programming language and the Flutter framework [8]. The research evaluates tasks of varying complexity, ranging from simple to challenging, and measures the approximate reduction in the developer's effort. These findings suggest that GitHub Copilot significantly enhances productivity, reducing the effort needed by approximately 70% for simple tasks and around 20% for more complex ones [8]. A study conducted by Moradi Dakhel et al. investigates the impact of GitHub Copilot on developer productivity, specifically in creating an HTTP server in JavaScript [9]. This research involved 95 professional programmers recruited through Upwork, capturing a diverse sample of experienced developers. These findings indicate that software developers utilizing AI assistance, such as GitHub Copilot, completed their tasks approximately 55.8% faster than those without AI support. This significant improvement highlights the potential of Copilot to streamline development processes and enhance efficiency in real world programming scenarios.

Nguyen and Nadi [10] conducted an empirical study on GitHub Copilot's effectiveness in generating code suggestions using LeetCode programming questions. They evaluated its performance across multiple programming languages, including Python, Java, JavaScript, and C, highlighting its capability to produce clear and low complexity solutions. However, they also identified limitations, such as the generation of suboptimal code and reliance on undefined helper methods. Mastropaolo et al. [11] investigated the robustness of deep learning-based code recommendation systems, including GitHub Copilot, using a dataset of 892 Java methods. Their findings revealed that semantically equivalent descriptions resulted in different code generation outcomes approximately 46% of the time. Similarly, Yetistiren et al. [12] assessed the code quality of GitHub Copilot, Amazon CodeWhisperer, and ChatGPT in the context of Python programming. Their evaluation found that ChatGPT outperformed the other tools in generating correct code solutions. They also emphasized the importance of input quality, demonstrating that well defined problem descriptions play a key role in successful code generation. Mehmood et al. [13] examined GitHub Copilot's ability to generate test cases, comparing them to manually written ones. Their analysis highlighted that AI generated test cases demonstrated comparable quality and effectiveness.

However, their study was limited to Python and a small set of files.

Sobania et al. [14] compared GitHub Copilot with Genetic Programming (GP) for program synthesis in Python. Their findings suggest that GP is better suited for tasks requiring numerous input/output examples, whereas Copilot performs well for problems defined through textual descriptions. Research on productivity and code quality has yielded mixed results regarding Copilot's impact on developer efficiency. While it can significantly boost productivity by generating large portions of code, the quality of its outputs often falls short, requiring extensive debugging. This underscores the need to balance productivity with code quality [15]. Researchers have observed a strong correlation between Copilot's acceptance rate and perceived productivity but caution that this metric alone does not fully reflect the complexity of developer experiences [16]. While Copilot can improve overall programming efficiency, it does not always shorten task completion time, as AI generated code often requires additional debugging [17].

Mozannar et al. [18] analyzed data from interactions with GitHub Copilot, proposing a utility theoretic framework to optimize decisions on displaying or withholding suggestions. Based on data from 535 programmers, their study demonstrated that a substantial portion of suggestions likely to be rejected by developers could be avoided. They also highlighted the significance of considering a programmer's latent, unobserved state when determining when to present suggestions. Additionally, their findings revealed that using suggestion acceptance as a reward signal for guiding display decisions can lead to lower quality suggestions. Baralla et al. [19] examined GitHub Copilot's potential to enhance developer productivity, particularly in the context of smart contract development on the blockchain. Their study highlighted Copilot's capacity to generate functional code, improve efficiency, and assist in routine development tasks such as code generation, debugging, and testing [19]. The research found that Copilot excelled in generating code for simpler smart contracts and standard token implementations, contributing to accelerated development processes. However, its performance weakened with more complex contracts, particularly in handling intricate blockchain specific logic and advanced security considerations. This underscores the tool's effectiveness in expediting basic tasks while still requiring significant human oversight for more complicated scenarios, especially regarding security and efficiency.

Chatterjee et al. [20] reported significant productivity improvements following the adoption of GitHub Copilot at ANZ Bank. During a six-week experiment, the group using Copilot completed tasks 42.36% faster than the control group. Productivity improvements varied by skill level: beginners showed a 52.27% improvement, intermediates 41.6%, and advanced users 40.48%. These results suggest that Copilot notably enhanced efficiency in software engineering tasks. Bakal et al. [21] evaluated GitHub Copilot's impact on productivity at ZoomInfo with over 400 developers. The study found consistent acceptance rates (33% for suggestions, 20% for lines of code) and high developer satisfaction (72%). Copilot proved useful for tasks like boilerplate code generation but struggled with domain specific logic [21].

In addition, an internal study conducted by GitHub focused on evaluating the impact of GitHub Copilot on developer efficiency and satisfaction. In an experiment with 95 participants, developers using Copilot completed tasks 55% faster than those without it, demonstrating its effectiveness in boosting productivity. Furthermore, a survey of over 2,000 developers revealed that 88% felt more productive in their work, 77% agreed they spent less time searching for information, and 87% experienced less mental effort on repetitive tasks, reducing cognitive load. Additionally, 74% reported being able to focus on more satisfying work, as Copilot alleviated the burden of repetitive coding tasks. This analysis highlights how Copilot enhances speed and improves overall job satisfaction among developers [22].

Further, a study conducted by GitHub in partnership with Accenture explored how developers integrate the tool into their daily workflows [23]. This collaboration aimed to assess the impact of Copilot on developer productivity and code quality. The findings indicated that an increase in pull requests is a strong indicator of the value delivered; Accenture developers experienced an 8.69% increase in pull requests. Since each pull request must undergo code review, the pull request merge rate serves as an excellent measure of code quality from the perspective of maintainers and coworkers. Accenture observed a 15% increase in the pull request merge rate, indicating that as the volume of pull requests grew, so did the number that successfully passed code review. Moreover, there was an 84% increase in successful builds, suggesting that not only were more pull requests moving through the system, but they were also of higher quality, as evaluated by both human reviewers and test automation. Developers accepted approximately 30% of suggestions from GitHub Copilot, and 90% reported committing code recommended by Copilot. Additionally, 91% of developers stated that their teams merged pull requests containing code suggested by Copilot. This study demonstrated a strong adoption and growing influence within the developer community.

### *B. REFLECTIONS ON LITERATURE*

**Task Complexity Matters**: Studies suggest that the productivity benefits of Copilot are more noticeable in simpler tasks, whereas complex or domain specific challenges often require additional developer intervention to maintain code quality.

**Quality vs. Speed Trade off**: While accelerated code generation is frequently cited as a key advantage, research indicates that AI generated code often requires extra debugging and validation. This highlights an ongoing need to refine AI tools to better balance speed and code quality.

**Context Dependent Effectiveness**: The tool's impact varies based on the developer's experience and the specific coding context. Some studies suggest that beginners may experience greater relative productivity gains, though they might also rely more on suggestions that do not always align with best practices.

**Beyond Acceptance Rates**: Metrics such as suggestion acceptance rates and pull request merge rates provide useful insights, but the literature suggests they may not fully capture the balance between productivity gains and the subsequent efforts required to ensure code quality.

Overall, GitHub Copilot represents a significant advancement in AI assisted software development, offering the potential to enhance efficiency across a broad range of coding tasks. Its ability to generate code rapidly and facilitate learning makes it an invaluable tool in modern software engineering. However, while its benefits are evident, striking a balance between accelerated code generation and maintaining high quality, reliable software remains a key challenge.

## III. Security Concerns with GitHub Copilot

While GitHub Copilot provides numerous advantages in terms of productivity and code generation, it also raises important security concerns that need to be addressed. One major issue is the potential introduction of vulnerabilities, as Copilot may suggest code that includes known weaknesses if such patterns are prevalent in the training data. This can lead to the generation of insecure code, such as hardcoded credentials, improper input validation, or insufficient error handling. For instance, it might suggest embedding sensitive information like API keys or passwords directly into the code or producing code that lacks proper input sanitization, potentially resulting in SQL injection or cross site scripting (XSS) vulnerabilities. Furthermore, this may expose sensitive information, leading to potential legal or intellectual property (IP) issues.

### *A. UNDERSTANDING PRIOR RESEARCH IN CONTEXT*

To thoroughly assess the security implications of GitHub Copilot, we have conducted a review of relevant research papers, focusing on identified vulnerabilities and the risks associated with AI assisted development. In this section, we will explore potential vulnerabilities, present research evidence on security risks, and highlight broader concerns from the developer community and industry experts regarding the safe use of AI in software development.

The study by Pearce et al. focused on evaluating GitHub Copilot's code generation across 89 scenarios, covering 25 different Common Weakness Enumerations (CWEs), particularly high-risk ones from MITRE's "Top 25 Most Dangerous Software Weaknesses" list [24]. This research found that 44% of the code generated by Copilot contained security issues, highlighting significant concerns regarding the tool's output [24]. Baralla et al. [19] also examined GitHub Copilot from a security standpoint, emphasizing its limitations in consistently applying advanced security patterns and detecting vulnerabilities in smart contracts. While Copilot excels in generating basic security features for standard token implementations, it struggles with more complex blockchain specific security issues. Its vulnerability detection and automatic program repair (APR) capabilities are unreliable, often requiring multiple prompts to address all identified issues. This highlights the critical need for human oversight, as Copilot generated code may introduce inconsistencies or security flaws. Consequently, while Copilot can aid in speeding up the development process, developers are encouraged to integrate additional security measures and review the code thoroughly before deploying it in production environments [19].

Siddque et al. [25] introduced SecurityEval, a comprehensive dataset designed to evaluate the security of machine learning based code generation models. Comprising 130 diverse samples mapped to 75 different vulnerability types from the Common Weakness Enumeration, SecurityEval allows for an in-depth assessment of models like InCoder and GitHub Copilot. The results reveal that both models can generate vulnerable code in certain scenarios. With its thoroughness, SecurityEval offers a valuable benchmark for evaluating the security capabilities of other code generation models in future research [25]. Siddique et al. [26] investigated the presence of code smells and security vulnerabilities in the datasets used to train code generation models and examined whether these issues are reflected in the generated output. The study employed Pylint and Bandit to evaluate three different training sets and assess the output produced by an open source transformer-based model and GitHub Copilot. The results showed that code smells and security vulnerabilities in the training data were propagated into the generated code. These findings highlighted the need for further improvements in code generation techniques and emphasized the importance of carefully curating and scrutinizing the training data to mitigate such issues in the output.

Majdinasab et al. [27] replicated the study by Pearce et al. [24]. to assess security vulnerabilities in newer versions of GitHub Copilot. While AI powered code generation tools like Copilot and Amazon CodeWhisperer enhance developer productivity, concerns persist regarding the security of their generated code. The study analyzed Python code suggestions using CodeQL and found that the percentage of vulnerable code has decreased from 36.54% to 27.25%. Despite these improvements, the findings confirm that Copilot continues to generate insecure code, highlighting the need for ongoing enhancements in AI assisted coding security.

Given that this research was published in 2022, one needs to assess whether there have been any improvements in addressing these issues in the latest version of GitHub Copilot to draw informed conclusions. In another empirical study conducted by Fu et al., code snippets generated by GitHub Copilot from GitHub projects were analyzed [28]. The study revealed 452 generated snippets with a high likelihood of security vulnerabilities. Specifically, 32.8% of the Python and 24.5% of the JavaScript snippets exhibited security issues. These vulnerabilities spanned 38 distinct Common Weakness Enumeration (CWE) categories, including critical ones like CWE-330: Use of Insufficiently Random Values, CWE-78: OS Command Injection, and CWE-94: Improper Control of Code Generation. Notably, eight of these CWEs are listed in the 2023 CWE Top-25, underscoring the severity of the issues [28].

### *B. REFLECTIONS ON LITERATURE*

**Ensuring Training Data Quality:** From our literature study, we find that vulnerabilities in generated code may stem from issues within the training datasets. This suggests that more effective curation and filtering of training data could help improve security outcomes in AI assisted coding.

**The Role of Human Oversight:** Based on the literature, while Copilot significantly accelerates code generation, human oversight remains essential. Ensuring a careful review of outputs is vital for developers, especially in security sensitive contexts like smart contract development, to help mitigate potential risks.

**Ongoing Security Enhancements:** Research indicates that newer versions of Copilot appear to have reduced certain vulnerabilities. However, we believe that the persistent presence of security risks points to the need for continuous research and improvements to further strengthen AI assisted coding security.

**Effectiveness Based on Context:** From our literature study, it appears that Copilot's performance varies depending on task complexity and domain specific requirements. While it seems highly effective in generating boilerplate and routine code, we find that it may face challenges in scenarios that demand deeper security awareness and more nuanced decision making.

However, recognizing the security risks associated with AI generated code, GitHub introduced significant enhancements in 2023 to address these concerns. A key improvement was launching an AI based vulnerability prevention system designed to block insecure coding patterns in real time, making GitHub Copilot's suggestions more secure. This model targets explicitly common vulnerable coding patterns, such as hardcoded credentials, SQL injections, and path injections, thereby mitigating risks at the code generation stage itself. These developments represent GitHub's ongoing efforts to enhance the security of Copilot's output while maintaining its productivity benefits [29].

On the other hand, we believe it is important to highlight the potential risks of developers becoming overly reliant on Copilot's suggestions. There is a risk that developers might unintentionally accept sub optimal or insecure code without sufficient scrutiny, which could negatively impact overall code quality and security standards. While Copilot can expedite the coding process, it remains an AI model that may occasionally produce insecure or ineffective code patterns. Despite its capabilities, developers need to remain vigilant in reviewing and rigorously testing all generated code to ensure that Copilot's contributions align with the desired quality and security standards.

## IV. Best Practices

GitHub Copilot has introduced a transformative approach to code generation, but from our perspective, its integration into development workflows demands a thoughtful balance of its strengths and limitations. While the tool offers significant potential, we believe its effective use requires careful consideration of both its capabilities and its risks. Drawing from our analysis of Copilot's features, studies, and real-world impacts, we bring to the forefront practices that can foster secure and efficient usage. Developers can consider these practices based on their specific work environments and project contexts to maximize Copilot's benefits while mitigating any potential risks.

**Vigilant Code Review***:* Developers are encouraged to maintain a rigorous approach to code review when incorporating AI generated code. This involves carefully examining all Copilot generated code for accuracy, security, and alignment with project requirements, particularly in sensitive areas like authentication, data handling, and encryption. AI generated code can inadvertently introduce inefficiencies or vulnerabilities, so incorporating thorough peer review practices is essential. Engaging multiple perspectives in validating code suggestions helps identify potential issues early, ensuring that the code maintains high standards of quality and security throughout the development process. To further enhance this review process, the latest version of Copilot introduces advanced code review capabilities [30], integrated with GitHub to help users iterate, validate, and integrate review comments efficiently. This feature, however, is currently unavailable in the free version.

**Use Security Tools***:* It's worth considering the integration of automated security testing tools (if not already implemented), such as Static Application Security Testing (SAST) and Dynamic Application Security Testing (DAST), into the development workflow when utilizing GitHub Copilot. These tools can help identify and address vulnerabilities in the generated code, ensuring that potential security issues are detected early in the development process. Incorporating these tools can enhance the security of projects, adding an extra layer of protection against threats.

**Educate and Train***:* Comprehensive developer training can play a key role in helping enterprise or corporate organizations maximize the potential of GitHub Copilot. By fostering an understanding of best practices for writing secure code, developers can become more proficient at recognizing and addressing common vulnerabilities. Additionally, regularly broadcasting important Copilot enhancements can help keep the team informed about new features and improvements, ensuring they are well equipped to fully utilize the tool's capabilities.

**Maintain Transparency and Feedback***:* Establishing a feedback loop with GitHub is crucial for enhancing Copilot. By reporting issues and providing feedback on improvements, developers contribute to the ongoing refinement of the tool. Additionally, maintaining clear documentation on how Copilot is integrated into projects, including configurations, guidelines, and usage practices, ensures that team members can reference best practices and understand the tool's application within their specific context. This transparency helps foster a culture of continuous improvement and accountability, leading to high quality, secure code.

**Legal and Ethical Considerations***:* When using GitHub Copilot, developers need to be mindful of the legal and ethical implications of the generated code. Vigilance regarding IP and copyright issues is crucial to avoid potential infringement, including compliance with relevant licenses. The "Finding Matching Code" feature, when enabled, helps by providing references to the matching code along with the associated number and type of licenses [31]. Ethical usage also requires adherence to data privacy and security protocols. Special attention is necessary when using Copilot in contexts involving sensitive or confidential information, as this may lead to unintended data exposure and compromise security standards. Establishing and following clear guidelines and usage policies for sensitive projects can be highly beneficial. By following these practices, developers can effectively leverage Copilot while maintaining legal and ethical integrity.

## V. Future Work

The growth of GitHub Copilot is evident, as approximately 30-40% of organizations surveyed by Gartner actively encourage and promote the adoption of AI coding tools. Additionally, 29-49% of respondents across various markets reported that their organizations allow using these tools but provide limited encouragement. This highlights a significant opportunity for organizations to actively embrace the AI wave. As noted in the GitHub Blog, the ongoing integration of AI tools into software development teams reflects a growing trend that organizations can consider tapping into for enhanced productivity and innovation [32]. As GitHub Copilot and similar AI driven code generation tools continue to evolve, several areas present further development and research opportunities. In this section, we present our views on potential avenues for future work, including technological improvements and broader implications for the software development industry.

**Programming Coverage***:* GitHub Copilot currently supports a variety of programming languages, including C, C++, C#, Go, Java, JavaScript, Kotlin, PHP, Python, Ruby, Rust, Scala, and TypeScript [33]. However, the extent of support for each language can vary, depending on the volume and diversity of training data available for that particular language. Expanding the breadth and depth of programming language support in GitHub Copilot can enhance its versatility and value for a broader range of developers. Incorporating additional languages, frameworks, and emerging ones allows developers across various fields and specialties to benefit from AI driven code generation. By offering support for a broader range of languages and frameworks, Copilot can meet the varied needs of the software development community and enhance overall productivity.

**Expand IDE's support***:* Expanding GitHub Copilot's support across a broader range of Integrated Development Environments (IDEs) could significantly elevate the overall development experience. Currently, Copilot is available in popular IDEs such as Visual Studio Code, Eclipse, JetBrains, Azure Data Studio, Vim/NeoVim, Visual Studio, and Xcode [34]. Enhancing Copilot's integration with even more IDEs can streamline workflows and allow developers to leverage AI driven code suggestions more seamlessly within their preferred development environments. By facilitating smoother interactions between Copilot and other software development tools, teams can foster better collaboration, increase productivity, and ultimately improve both the efficiency of project completion and the quality of the code.

**AI Assisted Software Design***:* Expanding GitHub Copilot's capabilities to include AI assisted software design represents a significant opportunity for enhancing the development process. By offering suggestions for software design and architecture in addition to code generation, Copilot could provide valuable assistance in the early stages of development. This expansion may involve generating design patterns, architectural diagrams, and high-level system components, allowing developers to create more robust and well-structured applications from the outset. Such features could improve collaboration among team members and streamline the transition from design to implementation, ultimately contributing to higher quality software outcomes.

**Legal and Ethical Considerations**: Future developments of GitHub Copilot could benefit from addressing intellectual property concerns by implementing mechanisms that prevent the generation of code that infringes on copyrighted or proprietary material. This involves enhancing the model's ability to avoid generating code that resembles existing proprietary code, thereby improving compliance with legal standards and fostering greater trust among developers. Legal experts have raised important questions regarding the ethical use of AI generated code, necessitating ongoing dialogue and regulation within the industry. Additionally, investigating the ethical and social implications of widespread AI code generation could provide valuable insights into its long-term effects on the software development industry and the broader tech ecosystem.

**Transparency and Accountability***:* Future development could focus on enhancing clarity regarding how GitHub Copilot generates code. Developers can gain a better understanding of the tool by providing detailed explanations of its suggestions, including the reasoning and sources used to generate responses. This approach helps build user confidence and encourages responsible use among development teams.

**Academic and Industry Research***:* Academic and industry research plays a crucial role in understanding the impact of AI driven code generation tools, such as GitHub Copilot, Cursor AI, Amazon Code Whisperer, and Google Codey, on various aspects of software development. Existing studies on developer productivity, code quality, and team dynamics have already provided valuable insights into how these tools influence real world practices. However, further in-depth studies can expand on these findings, offering a deeper understanding of the long-term implications of these tools. Comprehensive studies and case analyses will help clarify the evolving relationship between developers and AI tools, providing a clearer understanding of their long-term impact.

**User Customization**: Empowering users to customize the tool to their preferences leads to more relevant and accurate suggestions. The current experimental prerelease version of copilot chat offers to switch between a few LLMs (GPT 4o, Claude 3.5 Sonnet, Gemini 2.0 Flash, o1, o3-mini) and the option to add workspaces file, which allows user customization [35]. Along with these options for fine tuning AI behavior, creating user profiles to define tone and subject matter preferences can be a great addition from a user personalization standpoint. The current "custom instruction feature" in GitHub Copilot allows users to set parameters such as tool usage, language, and style [36]. As this paper is being written, this feature is in preview and has the potential for further changes and improvements. Additionally, implementing a feature to save session preferences can ensure that future suggestions align with the user's style, ultimately enhancing overall accuracy.

**Standardization of Evaluation Metrics***:* Establishing standardized evaluation metrics and benchmarking practices across AI based code generation tools can serve as a means for comparing their performance and effectiveness. This standardization can facilitate a clearer understanding of each tool's strengths and weaknesses, enabling developers and organizations to make informed choices based on consistent criteria. One example we can quote is various LLM benchmarks [37] on evaluation, which provide metrics for assessing capabilities across different tasks.

**Code Generation Tools Evaluation Improvements**: Recent research has extensively evaluated GitHub Copilot and similar AI driven code generation tools [38-41]. Researchers have highlighted the significance of various metrics in assessing the performance and effectiveness of these tools, focusing on metrics such as code acceptance rate, correctness ratio, reproducibility, similarity, validity, accuracy, and security vulnerabilities. Additionally, conducting large scale code quality evaluations in real time environments, rather than the typically controlled settings, is important. These evaluations can consider a breadth of programming languages, including emerging and niche languages, and incorporate evaluation metrics focusing on coding standards, security vulnerabilities, and maintainability. Developers can gain a deeper understanding of Copilot's capabilities from these large-scale code quality evaluations. This knowledge will equip them with valuable insights into the tool's performance and effectiveness, enabling more informed decisions in their coding practices.

## VI. CONCLUSION

GitHub Copilot is a powerful tool that enhances productivity by automating routine coding tasks and enabling rapid prototyping. However, its integration into development workflows raises important considerations, particularly around security, intellectual property, and code quality. Based on a literature study, we present insights into the benefits and challenges of using Copilot, and to address these, we offer our perspective on best practices for integrating Copilot into development workflows, focusing on responsible AI adoption and addressing security, intellectual property, and code quality concerns. Additionally, we highlight future research directions and propose iterative improvements to enhance Copilot's capabilities while mitigating the associated risks and ensuring continuous adaptation to emerging challenges. As AI tools like Copilot continue to evolve, their role in software development is likely to expand, prompting the need for ongoing reflection and adaptation. The continuous evolution of these tools underscores the importance of sustained research and iterative improvements to address current limitations. Looking ahead, it is crucial to critically assess how Copilot integrates into development workflows, refining best practices that not only enhance productivity but also mitigate risks and uphold core principles of software quality.

**SURESH BABU NETTUR** has received his Master of Science (M.S) degree from the Birla Institute of Technology and Science (BITS), Pilani, Rajasthan, India, and his Bachelor of Technology degree in Computer Science and Engineering from Nagarjuna University, Guntur, India. With over two decades of expertise, he has established himself as a thought leader in artificial intelligence, deep learning, and machine learning solutions. His work spans the development of scalable and intelligent systems across industries such as healthcare, finance, telecom, and manufacturing. Suresh has been at the forefront of integrating AI-driven innovations into real-world applications, leveraging cutting-edge technologies such as OpenAI models, GitHub Copilot, and custom deep learning architectures to deliver transformative solutions. His contributions include designing and implementing advanced machine learning models, optimizing deep learning architectures for resource-constrained environments, and integrating AI solutions into software development and testing pipelines. With significant experience in cross-functional team leadership and managing onsite-offshore collaboration models, he has successfully delivered AI-powered applications across cloud platforms like AWS.

Suresh is passionate about applying AI to solve complex problems in healthcare and finance, including predictive analytics, automation, and intelligent decision-making systems. He is proficient in Agile methodologies, Test-Driven Development (TDD), and service-oriented architectures (SOA), ensuring seamless integration of AI and machine learning into software systems. As an advocate for innovative AI applications, Suresh is committed to advancing the field through sustainable and impactful solutions that redefine industry standards and improve quality of life.

**SHANTHI KARPURAPU** received the Bachelor of Technology degree in chemical engineering from Osmania University, Hyderabad, India and the Masters technology degree in chemical engineering from Institute of Chemical Technology, Mumbai, India.
She has over a decade of experience leading, designing, and developing test automation solutions for various platforms across healthcare, banking, and manufacturing industries using Agile and Waterfall methodologies. She is experienced in building reusable and extendable automation frameworks for web applications, REST, SOAP, and microservices. She is a strong follower of the shift-left testing approach, a certified AWS Cloud practitioner, and a machine learning specialist. She is passionate about utilizing AI-related technologies in software testing and the healthcare industry.

**UNNATI NETTUR** currently pursuing an undergraduate degree in Computer Science at Virginia Tech, Blacksburg, VA, USA. She possesses an avid curiosity about the constantly evolving field of technology and software development, with a particular interest in Artificial Intelligence. She is passionate about gaining experience in building innovative and creative solutions for current issues in the field of software engineering.

**LIKHIT SAGAR GAJJA** pursuing a Computer Science Bachelor's degree at BML Munjal University, Haryana, INDIA. He is evident in showing his passion for the dynamic field of technology and software development. His specific interests include Artificial Intelligence, Prompt Engineering, and Game Designing technologies, highlighting his dedication to obtaining hands-on experience and developing innovative solutions for real-time issues in software engineering.

**SRAVANTHY MYNENI** earned master of science in information technology and management from Illinois institute of technology, Chicago, Illinois in 2017 and Bachelor's degree in computer science in 2013. She is currently working as an engineer focused on data engineering and analysis. She has 8+ years of experience in designing, building and deploying data centric solutions using Agile and Waterfall methodologies. She is enthusiastic about data analysis, data engineering and AI application to provide solutions for real world problems.

**AKHIL DUSI** currently pursuing Masters of Information Sciences at University of Indiana Tech, Indiana, USA. He is a passionate researcher and developer with a diverse background in software development, cybersecurity, and emerging technologies. He has a proven ability to deliver innovative and practical solutions. His work includes developing web and mobile-based applications, conducting vulnerability assessments and penetration testing, and leveraging cloud platforms for efficient infrastructure management. He is certified in cybersecurity and machine learning, reflecting a strong commitment to continuous learning and staying at the forefront of technological advancements. His research interests focus on artificial intelligence, IoT, and secure system design, with a vision to drive impactful innovations.